\documentclass[8.5pt,twoside,twocolumn]{article}
\usepackage[super,sort&compress,comma]{natbib} 
\usepackage{mhchem}
\usepackage{times,mathptmx}	
\usepackage{sectsty}
\usepackage{balance} 

\usepackage{graphicx} 
\usepackage{lastpage}
\usepackage[format=plain,justification=raggedright,singlelinecheck=false,font=small,labelfont=bf,labelsep=space]{caption} 
\usepackage{fancyhdr}
\usepackage{bm}
\usepackage{amsmath}
\usepackage{epstopdf}
\usepackage{color}

\usepackage{ulem}
\begin{document}
\normalem

\thispagestyle{plain}
\fancypagestyle{plain}{
\fancyhead[L]{\includegraphics[height=8pt]{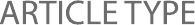}}
\fancyhead[C]{\hspace{-1cm}\includegraphics[height=20pt]{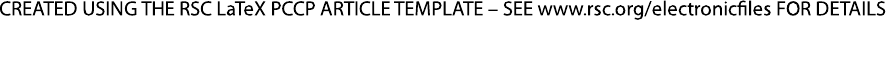}}
\fancyhead[R]{\includegraphics[height=10pt]{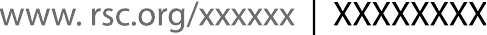}\vspace{-0.2cm}}
\renewcommand{\headrulewidth}{1pt}}
\renewcommand{\thefootnote}{\fnsymbol{footnote}}
\renewcommand\footnoterule{\vspace*{1pt}%
\hrule width 3.4in height 0.4pt \vspace*{5pt}} 
\setcounter{secnumdepth}{5}

\makeatletter 
\def\subsubsection{\@startsection{subsubsection}{3}{10pt}{-1.25ex plus -1ex minus -.1ex}{0ex plus 0ex}{\normalsize\bf}} 
\def\paragraph{\@startsection{paragraph}{4}{10pt}{-1.25ex plus -1ex minus -.1ex}{0ex plus 0ex}{\normalsize\textit}} 
\renewcommand\@biblabel[1]{#1}            
\renewcommand\@makefntext[1]%
{\noindent\makebox[0pt][r]{\@thefnmark\,}#1}
\makeatother 
\renewcommand{\figurename}{\small{Fig.}~}
\sectionfont{\large}
\subsectionfont{\normalsize}

\fancyfoot{}
\fancyfoot[LO,RE]{\vspace{-7pt}\includegraphics[height=9pt]{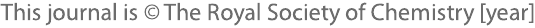}}
\fancyfoot[CO]{\vspace{-7.2pt}\hspace{12.2cm}\includegraphics{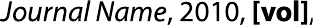}}
\fancyfoot[CE]{\vspace{-7.5pt}\hspace{-13.5cm}\includegraphics{RF}}
\fancyfoot[RO]{\footnotesize{\sffamily{1--\pageref{LastPage} ~\textbar  \hspace{2pt}\thepage}}}
\fancyfoot[LE]{\footnotesize{\sffamily{\thepage~\textbar\hspace{3.45cm} 1--\pageref{LastPage}}}}
\fancyhead{}
\renewcommand{\headrulewidth}{1pt} 
\renewcommand{\footrulewidth}{1pt}
\setlength{\arrayrulewidth}{1pt}
\setlength{\columnsep}{6.5mm}
\setlength\bibsep{1pt}

\newcommand{\citeN}[1]{\bibpunct{}{}{,}{n}{}{}\cite{#1}\bibpunct{}{}{,}{s}{}{}}
\definecolor{GB0}{rgb}{0.0,0.6,0.0} 
\definecolor{GB1}{rgb}{0.0,0.4,0.4} 
\definecolor{GB2}{rgb}{0.0,0.4,0.6} 
\definecolor{GB3}{rgb}{0.0,0.2,0.8} 
\definecolor{BR1}{rgb}{0.2,0.0,0.8} 
\definecolor{BR2}{rgb}{0.6,0.0,0.4} 
\definecolor{BR3}{rgb}{0.4,0.0,0.6} 
\definecolor{BR4}{rgb}{0.8,0.0,0.2} 
\newcommand{\changeA}[1]{#1} 
\newcommand{\changeB}[1]{#1} 
\newcommand{\changeC}[1]{#1} 
\newcommand{\changeD}[1]{#1} 
\newcommand{\changeE}[1]{#1} 
\newcommand{\changeF}[1]{#1} 
\newcommand{\com}[1]{} 

\twocolumn[
  \begin{@twocolumnfalse}
\noindent\LARGE{\textbf{Cooperative Self-Propulsion of Active and Passive Rotors $^\dag$}}
\vspace{0.6cm}

\noindent\large{\textbf{Yaouen Fily,\textit{$^{\ast a}$} Aparna Baskaran,\textit{$^{b}$} and
M. Cristina Marchetti\textit{$^{a}$}}}\vspace{0.5cm}

\noindent\textit{\small{\textbf{Received Xth XXXXXXXXXX 20XX, Accepted Xth XXXXXXXXX 20XX\newline
First published on the web Xth XXXXXXXXXX 200X}}}

\noindent \textbf{\small{DOI: 10.1039/b000000x}}
\vspace{0.6cm}

\noindent \normalsize{Using minimal models for  low Reynolds number passive and active rotors in a fluid, we characterize the hydrodynamic interactions among rotors and the resulting dynamics of a pair of interacting rotors. This allows us to treat in a common framework passive or externally driven rotors, such as magnetic colloids driven by a rotating magnetic field, and active or internally driven rotors, such as  sperm cells confined at boundaries. The hydrodynamic interaction of passive rotors \changeA{is known to} contain an azimuthal component $\sim 1/r^2$ to dipolar order that can yield the recently discovered ``cooperative self-propulsion" of a pair of rotors of opposite vorticity. While this interaction is identically zero for active rotors as a consequence of torque balance, we show that a $\sim1/r^4$ azimuthal component of the interaction arises in active systems to octupolar order. Cooperative self-propulsion, although weaker, can therefore also occur for pairs of active rotors.}
\vspace{0.5cm}
 \end{@twocolumnfalse}
  ]

\footnotetext{\dag~Electronic Supplementary Information (ESI) available.
}


\footnotetext{\textit{$^{a}$~Physics Department, Syracuse University, Syracuse, NY13244, USA; E-mail: yffily@syr.edu}}
\footnotetext{\textit{$^{b}$~Martin A. Fisher School of Physics, Brandeis University, Waltham, MA 02454, USA.}}



\section{Introduction}

There has been a lot of interest in the properties of individually propelled particles that take up energy from their environment and collectively generate motion and mechanical stresses \cite{Toner2005}. These systems exhibit remarkable collective behavior, quite different from their equilibrium counterparts.
One important class of propelled particles that has received relatively little attention is ``rotors'', that is particles that rotate in a fluid in response to externally applied or internally generated torques.
An 
external torque is provided for instance by a rotating magnetic field~\cite{Grzybowski2000,Grzybowski2001,Climent2007,Barbic2001} applied 
to particles carrying a magnetic moment, 
or by an optical trap~\cite{Friese1998,Friese1998a}.
Internally generated torques, on the other hand, originate from forces exerted by the rotor itself on the fluid.
We will refer to the first class of rotors driven by external torques as {\em passive} rotors, while the name of {\em active} rotors will be reserved to those that are internally driven. 
Several examples of active rotors are found in the living world, including sperm cells~\cite{Riedel2005,Friedrich2010,Friedrich2008,Gadelha2010,Watari2010}, bacteria~\cite{DiLuzio2005,Lauga2006,Lemelle2010,Giacche2010} and algae~\cite{Drescher2009} near a solid surface.
Various artificial swimmers, inspired by their living counterparts, have also been engineered over the past decade, and  provide realizations of active rotors~\cite{Dhar2006,Dreyfus2005,Walther2008,Wang2009} .
Other examples of internally driven or active rotors include the rotating motors found at the basal bodies of cilia and flagella anchored at the cell membrane~\cite{Brennen1977} and ATPase molecular motors embedded in fluid membranes~\cite{Boyer1997}.
In some cases, it is also possible to apply an additional external drive to an active swimmer, as with magnetotactic bacteria in a rotating magnetic field \cite{Erglis2007}.

Rotors generally move through a viscous fluid and  generate flow in the fluid.
This flow in turn affects the dynamics of other nearby rotors.
The goal of this paper is to characterize the flow field induced by various types of rotors and the resulting hydrodynamic interactions among them, 
\changeB{highlighting the differences between active and passive rotors. We then discuss how the flow field controls the rotors' dynamics.}
We limit our analysis to the case of rotors confined to a two-dimensional plane with their spinning direction perpendicular to that plane.
This constraint is often realized in experiments by confining rotors  to a liquid/air interface~\cite{Grzybowski2000,Grzybowski2001,Climent2007,Barbic2001},
or a thin layer of viscous fluid  embedded in a different unbounded viscous fluid~\cite{Lenz2003,Lenz2004,Leoni2010a}, or in the case of many swimming unicellular organisms simply because they are attracted to a boundary that in turn triggers the rotational dynamics~\cite{DiLuzio2005,Lauga2006,Lemelle2010,Giacche2010,Riedel2005,Friedrich2010}.

The main difference between passive and active rotors resides in the nature of the azimuthal component of the flow they create.
Passive rotors transfer external torque to the fluid, generating a dipolar
azimuthal flow that decays as $1/r^2$, with $r$ the distance from the rotor.
Active rotors do not transfer any net torque to the fluid, as the torque transmitted from the rotors to the fluid \changeB{is} balanced by an equal and opposite torque transmitted by the fluid to the rotors. As a result, the $1/r^2$  dipolar azimuthal component of the flow field is zero for active rotors.
 \changeB{This important distinction has been known for some time and was first pointed out in a classic paper by Batchelor~\cite{Batchelor1970}. More recently, it has been stressed by Lenz and collaborators~\cite{Lenz2003,Lenz2004} and by Michelin and collaborators~\cite{Michelin2010}, but is still sometimes overlooked in the literature.
 Our work shows that while the $1/r^2$ azimuthal component to the flow field vanishes as expected for active rotors, the cycle-averaged flow field of these force-free swimmers does contain a non-zero azimuthal in-plane component, although  only to octupole order in the multipole expansion. The corresponding contribution to the flow field hence decays as $1/r^4$ rather than as $1/r^2$ as for passive rotors. 
 
 This finding is an important new result of our work because
the existence of a finite azimuthal interaction alters significantly the collective dynamics of the rotors.
 It is known that} an azimuthal interaction between two \changeB{passive} rotors of the \emph{same vorticity} (\emph{i.e.}, spinning in the same direction) results in the pair revolving around each other along a circular path, as seen in experiments on externally driven rotors~\cite{Grzybowski2000,Grzybowski2001,Climent2007}. \changeB{It was shown recently by Leoni and Liverpool~\cite{Leoni2010a} that for  two rotors of \emph{opposite vorticity} the same $ 1/r^2$ azimuthal interaction turns the rotor pair into a self-propelled unit, in that the two rotors push each other in a common direction. The assumption of a $1/r^2$ azimuthal interaction made in Ref.~\citeN{Leoni2010a}  seems to restrict the result to the case of externally driven rotors, although much of the discussion in that paper refers to active rotor. Our work demonstrates that this type of ``cooperative self-propulsion'' also occurs for active rotors and arises not from a $\sim 1/r^2$ azimuthal coupling, that vanishes for  force-free particles, but from the new $1/r^4$ octupolar contribution to the flow that we have identified. We also show that for active rotors the additional  attractive and repulsive interactions arising among pullers and pushers from the radial dipolar component of the flow field can lead to the formation of a bound self-propelled pair consisting of two oppositely rotating swimmers.

To highlight the difference between passive and active rotors, in the following we first review known results for the simplest passive rotor consisting of  a sphere driven by an external torque. Then we analyze the flow field and hydrodynamic interactions of a minimal model for an active rotor, consisting a one-bead swimmer  that rotates by pushing on the fluid with a force oriented at a finite angle to the swimmer's long axis (see Fig.~\ref{fig_swimmer1}).}

\section{Model}
\label{model}

We consider rotors moving through a viscous incompressible fluid in the $xy$ plane and spinning around the $z$ axis.
Due to the small size of the rotors, inertial effects are negligible both on the dynamics of the fluid and that of the rotors.
The fluid is therefore described in the zero Reynolds number limit where the flow field ${\bf u}$ is solution of the incompressible Stokes equation
\begin{subequations}\label{incStokes}
\begin{gather}
\eta \nabla^2{\bf u} -\nabla p=-{\bf f}({\bf r})\;,
\label{stokes} \\
\nabla\cdot{\bf u}=0\;,
\label{incompressible}
\end{gather}
\end{subequations}
where $\eta$ is the viscosity of the fluid and ${\bf f}$ is the force density exerted by the rotors on the fluid.
The solution of Eqs.~\eqref{incStokes} is given by
\begin{equation}
u_i({\bf r})=\frac{1}{8\pi\eta}\int d{\bf r}'{\mathcal O}_{ij}({\bf r}-{\bf r}')f_j({\bf r}')\;,
\label{stokes-solution}
\end{equation}
where ${\cal O}_{ij}({\bf r})$ is the Green's function of Eqs.~\eqref{incStokes} and depends on the geometry of the problem.
In the following we use  the \emph{Oseen tensor} given by
${\cal O}_{ij}({\bf r})=\frac{1}{r}(\delta_{ij}+\hat{r}_i\hat{r}_j)$ with $\hat{r}_i=r_i/r$,
which corresponds to an unbounded fluid in three-dimensions,
but can also describe the flow field generated by particles confined to a fluid-air interface \cite{Cichocki2004,Cichocki2004a} in the plane of the interface. Our results are, however, easily extended to other Green's function describing different physical situations (see section~\ref{discussion}).

The rotors dynamics is controlled by force and torque balance, given by
\begin{subequations} \label{FTbal}
\begin{gather}
{\bf F}^d+{\bf F}^a={\bf 0} \;,
\label{FB} \\
{\bm \tau}^d+{\bm \tau}^a+{\bm \tau}^e={\bf 0} \;,
\label{TB}
\end{gather}
\end{subequations}
where the superscripts $d$, $a$ and $e$ stand respectively for \emph{drag}, \emph{active} and \emph{external}, and the forces and torques are exerted on the rotors.
We only consider below situations where the external force is  zero (in particular, the rotors are assumed to be at neutral buoyancy), but generally allow for the presence of an external torque.

\subsection{Single passive rotor}
\label{passive1}

We model a passive rotor as a single neutrally buoyant sphere of radius $a$ rotating under the action of an externally imposed torque, ${\bm \tau}^e=\tau^e \hat{\bf z}$.
In this case there are no active forces or torques, and
the drag force and torque exerted by the fluid on the sphere are given by
\changeA{
\begin{subequations} \label{drag}
\begin{gather}
{\bf F}^d=-\zeta
\left( {\bf v}_\textsc{c}
-\left[ \left( 1+\frac{a^2}{6}\nabla^2 \right)
 {\bf u}^0 \right]_{{\bf r}_\textsc{c}}
\right)
\label{drag_F} \\
{\bm \tau}^d=-\zeta^\textsc{r} \left( {\bm \omega}
-\frac{1}{2} \left[ \nabla\times{\bf u}^0 \right]_{{\bf r}_\textsc{c}}
\right) \;,
\label{drag_T}
\end{gather}
\end{subequations}
}
where $\zeta=6\pi\eta a$ and $\zeta^\textsc{r}=8\pi\eta a^3$ are the translational and rotational friction of the sphere, 
${\bf r}_\textsc{c}$ is the position of its center,
${\bf v}_\textsc{c}=d{\bf r}_\textsc{c}/dt$ is its velocity and 
${\bm \omega}$ is its angular velocity describing rotations about the center.
The flow ${\bf u}^0({\bf r})$ is the one that would be obtained if the sphere was removed and replaced with fluid. 
\changeB{The %
\changeF{laplacian term on the right hand side of Eq.~\eqref{drag_F} is} the Fax\'en law correction for the friction on spherical objects in a viscous fluid and is important when flow gradients occur on length scales comparable to the sphere's diameter.}~\cite{Pozrikidis1997}
The translational and rotational velocity of the sphere are immediately obtained as
\changeA{
\begin{subequations}\label{dyn_passive1}
\begin{gather}
{\bf v}_\textsc{c} =
\left[ \left( 1+\frac{a^2}{6}\nabla^2 \right)
 {\bf u}^0 \right]_{{\bf r}_\textsc{c}}
\;,
\label{FB_passive1} \\
{\bm \omega} = \frac{\tau^e}{\zeta^\textsc{r}}\hat{\bf z}
+\frac{1}{2} \left[ \nabla\times{\bf u}^0 \right]_{{\bf r}_\textsc{c}}
\;.
\label{TB_passive1}
\end{gather}
\end{subequations}
}
\changeB{If there is no externally imposed flow and only one rotor is present in the fluid, then ${\bf u}^0({\bf r})=0$.
In this case a single passive rotor} does not translate (${\bf v}_\textsc{c}=0$) and it  rotates around the $z$ axis with angular velocity ${\bm \omega}=\hat{\bf z}\tau^e /\zeta^\textsc{r}$.

\begin{figure}[h]
\centering
\includegraphics[width=0.85\linewidth]{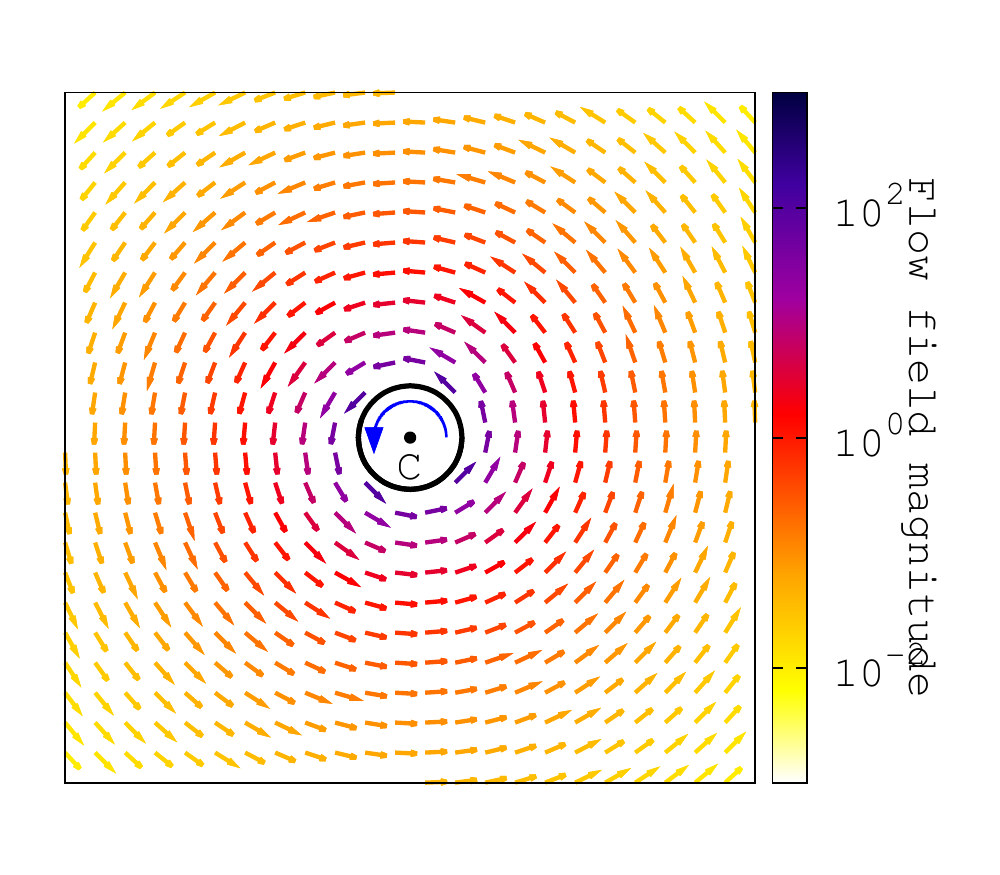}
\caption{ (color online)
Map of the flow field generated in the $xy$ plane by a passive rotor modeled as a sphere driven by an external torque (represented by a circular arrow inside the rotor) directed along the out-of-plane direction $z$. The rotation is counter-clockwise and occurs around the $z$ axis.
The fluid velocity is shown as arrows with color coded magnitude.
}
\label{fig_passive1}
\end{figure}

The flow field due to the rotation of \changeF{the} sphere is obtained by enforcing a no-slip boundary condition at its rotating surface \changeB{and is given by}
~\cite{Pozrikidis1997}
\begin{align}
{\bf u}({\bf r})=\frac{\tau^e}{8\pi\eta r^2} \hat{\bf z}\times\hat{\bf r} \;.
\label{flowPassive1}
\end{align}
The flow field is purely azimuthal in the $xy$ plane and it decays as $1/r^2$ (see Fig.~\ref{fig_passive1}). 

It is important to note that this azimuthal flow field is a general property of passive rotors. 
This 
can be seen by expanding Eq.~\eqref{stokes-solution} in multipoles.
The monopole term vanishes because there are no external forces acting on the rotor. The first nonzero term in the multipole expansion is the dipolar term that has the form
\begin{multline}
u_i({\bf r})=-\frac{1}{8\pi\eta} 
[{\cal H}_{ijk}({\bf r}) + {\cal G}_{ijk}({\bf r}) ] \int d{\bf r'} r'_k f_j({\bf r'})
\;,
\label{multexp1}
\end{multline}
where ${\cal H}_{ijk}=(\delta_{jk}-3\hat{r}_j\hat{r}_k) \hat{r}_i/r^2$ and
${\cal G}_{ijk}=\epsilon_{inq} \epsilon_{njk} \hat{r}_q/r^2$
are the $jk$ symmetric and antisymmetric parts of $\partial_k {\cal O}_{ij}$.
The decomposition of the flow into its ${\cal H}_{ijk}$ part (known as a \emph{stresslet}) and ${\cal G}_{ijk}$ part (known as a \emph{rotlet}) coincides with a polar decomposition in the plane perpendicular to the axis of rotation: the \emph{stresslet} is purely radial while the \emph{rotlet} is purely azimuthal.
The antisymmetric part of the integral represents the total torque exerted by the rotor on the fluid.
Torque balance requires that this equals the external torque applied to the rotor, or
\begin{equation}
\epsilon_{ikj} \int d{\bf r'} r'_k f_j({\bf r'})=\tau^e_i\;.
\end{equation}
As a result, the antisymmetric or \emph{rotlet} part of the flow field given in Eq.~\eqref{multexp1}  equals the azimuthal flow field given in Eq.~\eqref{flowPassive1} for all passive rotors and vanishes when $\bm\tau^e=0$.
In other words, the azimuthal flow field of an externally driven rotor is to leading (dipolar) order identical to that of a sphere driven by the same torque, regardless of the details of the rotor.
The symmetric part of the integral, on the other hand, is not constrained and depends on the geometry of the rotor.

\subsection{Single active rotor}
\label{swimmer1}

\begin{figure}[h]
\centering
\includegraphics[width=0.9\linewidth]{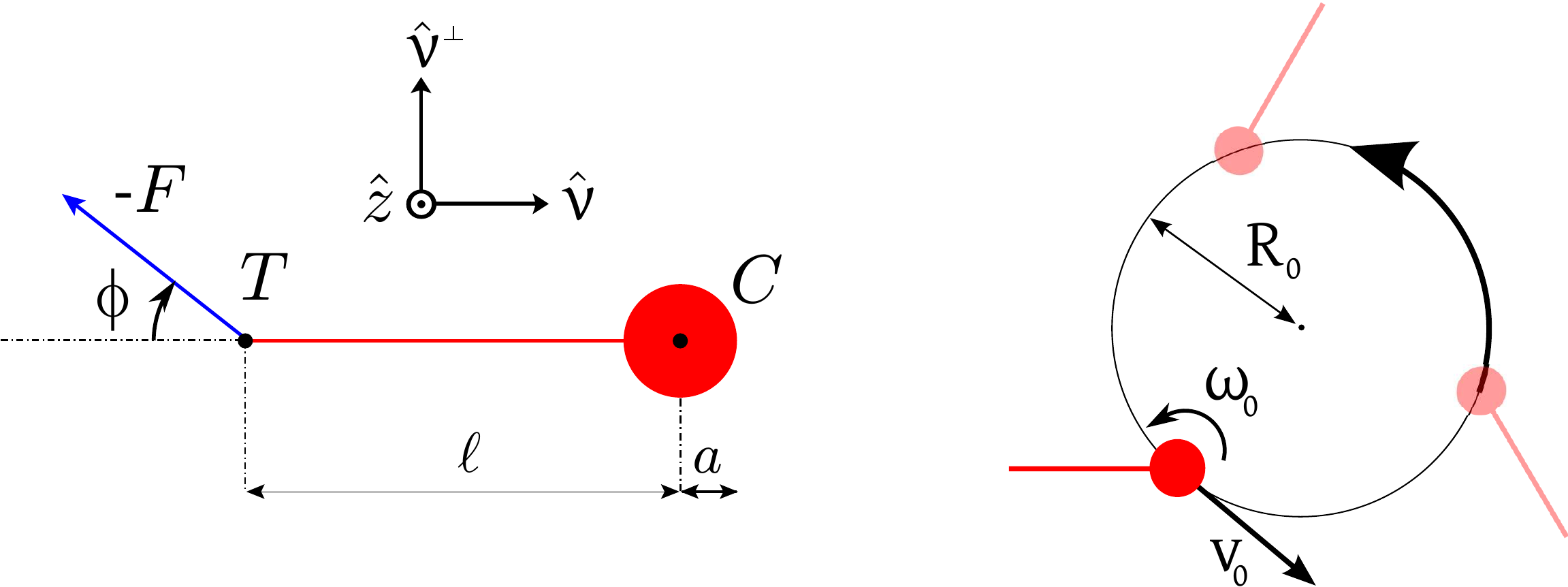}
\caption{
Left: Schematic of the single-bead swimmer.
The swimmer exerts an active force $-{\bf F}$ on the fluid applied at point T at an angle $\phi$ to the swimmer's axis defined by the unit vector $\hat{\bm \nu}$. The bead has radius $a$ and its center $C$ is the hydrodynamic center of the swimmer.
The rotation is counterclockwise about the $z$ axis normal to the page through $C$.
Right: Schematic of the swimmer's motion. The swimmer, drawn at three different times, undergoes a full rotation around itself every $T_0=2\pi/\omega_0$ while translating with velocity $\text{v}_0$, resulting in a circular trajectory of radius $R_0=\text{v}_0/\omega_0$.
}
\label{fig_swimmer1}
\end{figure}

An active swimmer is one that propels itself by pushing on the fluid.
In order to obtain net propulsion  in the absence of an externally applied  force or torque, the drag center $C$ and the thrust center $T$ of the swimmer must be located at two different points. With this in mind, we consider the simplest realization of a rotating swimmer, consisting of
a spherical bead of radius $a$ centered at point $C$ and providing the drag and a point force $-{\bf F}$ exerted on the fluid at the end $T$ of a fictitious ``flagellum" of length $\ell$ providing the thrust.
We denote by $\hat{\bm \nu}$ a unit vector pointing along the swimmer's long axis (chosen as the direction of the flagellum), and by $\phi$ the angle between $\hat{\bm \nu}$ and ${\bf F}$ (see Fig.~\ref{fig_swimmer1}).
The dynamics of the swimmer in the absence of external torque is governed by Eqs.~\eqref{FTbal} with ${\bm \tau}^e=0$ and ${\bf F}^a={\bf F}$, while
the active torque ${\bm \tau}^a=-\ell\hat{\bm \nu}\times{\bf F}$ is equal to the moment in $C$ of ${\bf F}^a$.
The drag force and torque are given by Eqs.~\eqref{drag}.
\changeA{%
For an isolated swimmer, ${\bf u}^0$ reduces to the flow field 
created by the thrust force $-{\bf F}$ exerted on the fluid at point $T$.
The linear and angular velocities of the swimmer are then given by
\begin{subequations} \label{dyn_swim1}
\begin{gather}
\text{v}_{\textsc{c}i}
 =\frac{F_j}{\zeta} \left[
\delta_{ij}
-\frac{3a}{4\ell}\left( \delta_{ij} +\hat{\nu}_i\hat{\nu}_j\right)
-\frac{a^3}{4\ell^3}\left( \delta_{ij} -3\hat{\nu}_i\hat{\nu}_j\right)
\right]
\;, \\
{\bm \omega}
 =-\frac{\ell}{\zeta^\textsc{r}}\hat{\bm \nu}\times{\bf F}
\left( 1-\frac{a^3}{\ell^3} \right)
\;.
\label{dynO_swim1}
\end{gather}
\end{subequations}
The thrust force $-{\bf F}$ exerted by the swimmer on the fluid at $T$ affects the dynamics in two ways:
(i) ``directly'' through the reaction force $+{\bf F}$ exerted back on the swimmer by the fluid at $T$, 
and (ii) ``indirectly'' through ${\bf u}^0$.
These two contributions cancel each other when $\ell= a$, i.e., when the thrust force is exerted at the surface of the bead, giving ${\bf v}_\textsc{c}={\bf 0}$ and ${\bm \omega}={\bf 0}$ in this limit.
Conversely, the ``direct'' contribution is dominant when $a\ll\ell$.
Since $\hat{\bm \nu}$ and ${\bf F}$ both lie in the $xy$ plane, ${\bm \omega}=\omega_0{\bf \hat{z}}$ is along $z$. An isolated swimmer then translates at ${\bf v}_\textsc{c}$ while rotating at an angular  frequency $\omega_0$ about an axis  through its center $C$ directed along the $z$ direction. The trajectory of $C$ is a circle of radius
$R_0=\text{v}_0/\omega_0$ in the $xy$ plane, as shown on Fig.~\ref{fig_swimmer1}, where $\text{v}_0=\vert{\bf v}_\textsc{c}\vert$. The period of rotation is $T_0=2\pi/\omega_0$.
Note that when $\phi\rightarrow0$, $R_0$ diverges and the trajectory becomes a straight line, as expected for a linear swimmer exerting a trust force directed  along its long axis.

In order to keep the model tractable, we work in the $a\ll\ell$ limit.
Since $R_0/\ell\sim{\cal O}\left((a/\ell)^2\right) /\sin\phi$, the radius of the circle is then negligible for all finite values of $\phi$ and the swimmer is merely rotating rigidly about its center~$^\dag$.
Moreover, the dynamics is dominated by the ``direct'' contribution of the thrust force and the velocity and angular velocity are simply given by
${\bf v}_\textsc{c}\simeq{\bf F}/\zeta$ and ${\bm \omega}\simeq-\ell F\sin\phi \hat{\bf z}/\zeta^\textsc{r}$.
}
\footnotetext{\dag~%
\changeA{A detailed analysis of the role of a finite $a$, including the consequence of $R_0$ not being negligible, can be found in the Electronic Supplementary Information.}
}

The flow field created by the one-bead swimmer is the sum of the flows produced by three sources:
a point force $-{\bf F}$  located at $T$ corresponding to thrust, the translation of the bead with velocity ${\bf v}={\bf F}/\zeta$, and the rotation of the same bead with angular velocity ${\bm \omega}=-\ell\hat{\bm \nu}\times{\bf F}/\zeta^\textsc{r}$.
This gives

\begin{multline}
u_i({\bf r})
=
-\frac{1}{8\pi\eta} {\cal O}_{ij}({\bf r}-{\bf r}_\textsc{c}+\ell\hat{\bm \nu}) F_j
+\frac{1}{8\pi\eta} {\cal O}_{ij}({\bf r}-{\bf r}_\textsc{c}) F_j \\
+\frac{\ell}{8\pi\eta} {\cal G}_{ijk}({\bf r}-{\bf r}_\textsc{c}) F_j \hat{\nu}_k\;,
\label{flowSwimmer1}
\end{multline}
where we have used ${\bf r}_\textsc{t}={\bf r}_\textsc{c}-\ell\hat{\bm \nu}$
and expressed ${\bf v}$ and ${\bm \omega}$ in terms of $\ell$, $\hat{\bm \nu}$ and ${\bf F}$.
In order to study the the far field behavior of the flow, we expand Eq.~\eqref{flowSwimmer1} in powers of $\ell/r$. The flow field can then be written in the form of a multipole expansion, as
\begin{equation}
{\bf u}({\bf r})=\sum_{n=0}^\infty{\bf u}^{(n)}({\bf r})\;,
\end{equation}
where ${\bf u}^{(n)}({\bf r})\sim (F/\eta r)(\ell/r)^n$ represents the $(n+1)$-pole contribution to the flow.
In the absence of external forces the monopole term ${\bf u}^{(0)}({\bf r})$ vanishes.
The lowest non-vanishing term of this expansion is the dipolar one. 
We note, however, that as a consequence of torque balance the \emph{rotlet} due to the rotation of the bead, given by the last term on the right hand side of  Eq.~\eqref{flowSwimmer1}, is precisely canceled by the \emph{rotlet} component of the first two terms, representing the flow due to the force dipole formed by the two point forces.
As a result, only the \emph{stresslet} (radial) part of the flow is nonzero to dipolar order, with
\changeE{
\begin{align}
u_i^{(1)}({\bf r}) 
& =-\frac{1}{8\pi\eta} 
\ell F_j \hat{\nu}_k {\cal H}_{ijk} \nonumber \\
& = \frac{\ell F}{16\pi\eta} 
\Big[ \cos\phi +3\cos(2\theta+\phi)\Big] 
\frac{\hat{r}_i}{r^2}
\;,
\end{align}%
where $\theta$ is the angle between $\hat{\bm \nu}$ and ${\bf r}$.
}%
As we pointed out in section~\ref{passive1}, this is a general consequence of torque balance and holds for any rotating swimmer.

Retaining terms up to  third (octupolar) order in the expansion, the total flow field of a single  one-bead swimmer is given by
\begin{multline}
u_i({\bf r})
= -\frac{1}{8\pi\eta} \Big[
\ell \hat{\nu}_k {\cal H}_{ijk}
+\frac{\ell^2}{2} \hat{\nu}_k \hat{\nu}_l \partial_{kl} {\cal O}_{ij} \\
+\frac{\ell^3}{6} \hat{\nu}_k \hat{\nu}_l \hat{\nu}_m \partial_{klm} {\cal O}_{ij}
\Big] F_j
+{\cal O}\left(\frac{1}{r^5}\right)
\;,
\label{flowSwimmer1Multi}
\end{multline}
where all the derivatives of the Oseen tensor are evaluated at $({\bf r}-{\bf r}_\textsc{c})$.

\begin{figure}[h]
\centering
\includegraphics[width=0.85\linewidth]{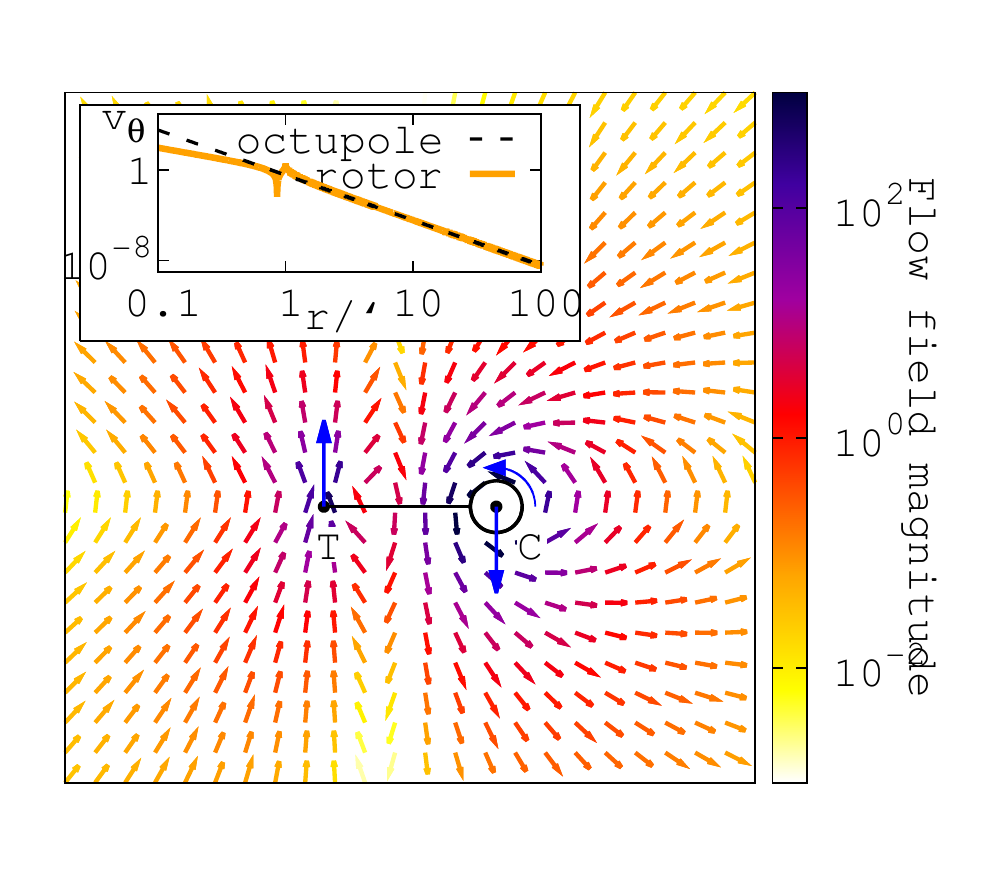}
\caption{
Map of the instanteneous flow field generated in the $xy$ plane by a one-bead pure active rotor (\emph{i.e.} thrust force perpendicular to the main axis) rotating counter-clockwise around the out-of-plane direction $z$.
The fluid velocity is shown as arrows with color coded magnitude.
The arrows represent the forces (straight arrows) and torque (curved arrow) exerted by the rotor on the fluid.
The inset shows the flow field averaged over a rotation. The averaged flow field is azimuthal and dominated by the average flow field of a point octupole. Distance is normalized by the length $\ell$ of the rotor.}
\label{fig_active1}
\end{figure}

On time scales longer that the period of rotation $T_0$ of the swimmer(see Fig.~\ref{fig_swimmer1}), it is useful to consider the cycle-averaged flow field, denoted by $\langle\ ...\rangle$ below. \changeB{As shown in Appendix~\ref{period_avg}, the average over a period can be obtained  by averaging the instantaneous flow over the orientation $\hat{\bm \nu}(t)$ of the swimmer's axis. Since the active force $F_j$ in Eq.~\eqref{flowSwimmer1Multi} can be written as 
$F_j=F\cos\phi\hat{\nu}_j-F\sin\phi\left({\bf \hat{z}}\times\bm{\hat \nu}\right)_j$, it is then evident that the dipolar and octupolar terms (first and third term on the left hand side of Eq.~\eqref{flowSwimmer1Multi}) \changeE{contain even powers of the} unit vector $\bm\hat{\nu}$, while the quadrupolar term is cubic in $\bm\hat{\nu}$. 
The latter will therefore vanish when we average over the direction of the swimmers' axis. This result is generic for swimmers \changeF{confined to a plane}. It should be stressed, however, that the \emph{instantaneous} quadrupolar flow field is of course nonzero at all times.} 

The cycle-averaged dipolar \emph{stresslet} contribution to the flow is isotropic and radial and is entirely controlled by the component $F\cos\phi$ of the propulsive force parallel to the axis of the dumbbell, with
\begin{align}
\left\langle {\bf u}^{(1)}({\bf r}) \right\rangle
= \frac{\ell F \cos\phi}{16\pi\eta r^2} \hat{\bf r}
\;.
\label{flowSwimmer1DipoleAvg}
\end{align}
The quadrupolar term in Eq.~\eqref{flowSwimmer1Multi} changes sign when the rotor is rotated by $\pi$ and thus does not contribute to the cycle-averaged flow.
The octupolar term gives rise to both a radial and an azimuthal contributions upon cycle averaging, with
\begin{align}
\left\langle {\bf u}^{(3)}({\bf r}) \right\rangle
= \frac{\ell^3 F}{16\pi\eta r^4} \left( \frac{3}{8}\cos\phi\; \hat{\bf r}
+\frac{3}{4}\sin\phi\; \hat{\bf z}\times\hat{\bf r} \right)
\;.
\label{flowSwimmer1OctupoleAvg}
\end{align}
The radial part \changeB{will be neglected below as it yields a correction to the cycle-averaged dipolar flow, with a faster decay.}
The azimuthal part, on the other hand, represents the leading order azimuthal flow. \changeB{Both Fax\'en effects and the finite value of the radius $R_0$ of the circular trajectory of the center of the rotor give corrections of order $(a/\ell)^2$ or higher to the coefficients of both the radial and azimuthal octupolar flow fields, but do not change their qualitative behavior and will be neglected here. The full expressions for the cycle averaged octupolar flow field can be found in the Electronic Supplementary Material$^\dag$.}

Keeping only the leading contribution of each component (\emph{i.e.} the radial dipolar term and the azimuthal octupolar term), \changeB{and neglecting ${\cal O}(a/\ell)$ corrections to all the coefficients}, the total rotation-averaged flow of a single one-bead swimmer is given by
\changeA{
\begin{multline}
\left\langle {\bf u}({\bf r}) \right\rangle = 
\frac{F\cos\phi}{16\pi\eta\ell}
\left[
\frac{\ell^2}{r^2}
+{\cal O}\left(\frac{\ell^4}{r^4}\right)
\right]
\cos\phi\ \hat{\bf r} \\
+ \frac{3F\sin\phi}{64\pi\eta\ell} ~\frac{\ell^4}{r^4}
 \hat{\bf z}\times\hat{\bf r}
+{\cal O}\left(\frac{\ell^5}{r^5}\right)
\;.
\label{eq:flowSwimmer1}
\end{multline}
}
The azimuthal term is proportional to the active torque $\ell{\bf F}\times\hat{\bm \nu}=-\ell F \sin \phi \hat{\bf z}$,
while the radial term is proportional to  ${\bf F}\cdot\hat{\bm \nu}=F\cos\phi$.
The latter is zero for a ``pure rotor'' (${\bf F}\cdot\hat{\bm \nu}=0$),
attractive for a ``puller'' (${\bf F}\cdot\hat{\bm \nu}<0$) and repulsive for a ``pusher'' (${\bf F}\cdot\hat{\bm \nu}>0$).

\section{Hydrodynamic interaction of two rotors}
\label{interactions}

In the presence of other rotors, the flow field generated by one rotor yields hydrodynamic interactions on the others. These interactions in turn control the collective rotor dynamics. To understand the role of such hydrodynamic couplings, in this section we consider  the effect of the fluid mediated interaction on the
dynamics of pairs of rotors. Again, we will use minimal models for both passive and active rotors, corresponding, respectively, to a single rotating sphere and to the  one-bead swimmer discussed in the previous section.

We consider two rotors in the $xy$ plane. The  positions of their drag centers are denoted by
vectors $\left\{ {\bf r}_\alpha\right\}_{\alpha=1,2}$. In addition to the
fluid mediated hydrodynamic interactions, the two rotors are assumed to
experience a short-range central repulsive force
that opposes their overlapping.
The equations of motion of the two such rotors have the form
\changeA{
\begin{equation}
\partial_t{\bf r}_\alpha=
\left[ \left( 1 +\frac{a^2}{6}\nabla^2 \right)
{\bf u}_{\beta} \right]_{{\bf r}_\alpha}
-\frac{1}{\zeta}g(r_{\alpha\beta})\hat{\bf r}_{\alpha\beta}\;,  
\label{EoM}
\end{equation}
}
where ${\bf u}_\beta$ is the flow field induced by the rotor $\beta$ and 
$g(r_{\alpha\beta})$ is the magnitude of the repulsive central force 
and vanishes at distances larger than the effective diameter $d$ of the rotors,
with ${\bf r}_{\alpha\beta}={\bf r}_\alpha-{\bf r}_\beta$ $^\ddag$
\footnotetext{\ddag~When an explicit form was needed for the repulsion force (\emph{e.g.}, to run simulations, not shown), we used
$g(r)=\left[(r/d)^3-1\right]/r^2$ if $r<d$ and $g(r)=0$ if $r>d$.}.

\changeE{
For passive spheres, the flow field ${\bf u}$ that appears in Eq.~\eqref{EoM} is given by Eq.~\eqref{flowPassive1}.
For the active one-bead swimmer, we only consider the dynamics on time scales large compared to the period of rotation $T_0$.
\changeF{As shown in Appendix~\ref{period_avg}}, in this limit we can replace the instantaneous flow field ${\bf u}$ in Eq.~\eqref{EoM} by its cycle averaged value given by Eq.~\eqref{eq:flowSwimmer1}~$^\dag$.}%
\footnotetext{\dag~\changeD{If the radius $R_0\sim a^2/\ell$ of the trajectory of a single rotor were not negligible, 
the fact that both rotors are moving around a circle would have to be taken into account when performing the period average.
This case is treated in detail in section~2 of the Electronic Supplementary Information.}}
This can be written in a general form that applies to both passive (representing in this case the \emph{instantaneous} flow) and active rotors as%
\begin{equation}
\left\langle{\bf u}_\beta({\bf r}_\alpha)\right\rangle =
F^{a,p}(r_{\alpha\beta}) \hat{\bf r}_{\alpha\beta}
+A^{a,p}(r_{\alpha\beta})
\epsilon_{\beta}\hat{\bf z}\times \hat{\bf r}_{\alpha\beta}  \label{3.2}
\end{equation}
\changeE{
The first term is the radial flow, with 
$F^a=(\ell F \cos\phi)/(16\pi\eta r^2)$ and $F^p=0$.
The second term is the azimuthal flow of strength $A$, with
$A^p=\tau^e/(8\pi\eta r^2)$ and $A^a=(3\ell^3 F\sin\phi)/(64\pi\eta r^4)$.
A parameter $\epsilon_\alpha=\pm 1$ is introduced to
characterize the vorticity of the rotor, with  $\epsilon_\alpha=\pm1$ 
corresponding to clockwise (+1) and counter-clockwise (-1) rotations.%
} 

\changeA{
The term proportional to $a^2$ in Eq.~\eqref{EoM} contributes to the interaction to octupolar order through the laplacian of the dipolar flow field.
For a passive sphere, the dipolar flow is a rotlet and its laplacian is zero.
For a one-bead swimmer, the dipolar flow is given by Eq.~\ref{flowSwimmer1DipoleAvg} and its laplacian has the same structure as the radial part of ${\bf u}^{(3)}$, with an extra $(a/\ell)^2$ factor.
We can therefore neglect it as well.
}

To study the coupled dynamics of the rotor pair we introduce 
the center of "mass" and relative coordinates, ${\bf R}=({\bf r}_1+{\bf r}_2)/2$ and ${\bf r}={\bf r}_1-{\bf r}_2$, respectively.
Without loss of generality we assume that rotor $1$ is rotating clockwise $\epsilon_1=+1$. The equations of motion of the two rotors can be rewritten in the form
\begin{gather}
\partial_t{\bf R}=\frac{1}{2} A^{a,p}(r) \left( 1-\epsilon \right)
\hat{\bf z}\times\hat{\bf r}\;,  \label{3.4}\\
\partial_t{\bf r}=A^{a,p}(r) (1+\epsilon)
\hat{\bf z}\times\hat{\bf r}+2\left[ F^{a,p}(r)
-\frac{g(r)}{\zeta}\right] \hat{\bf r}\;,  \label{3.3}
\end{gather}
where $\epsilon=+1$ corresponds to "like" rotors, rotating in the same direction, while $\epsilon=-1$ describes "unlike" rotors, with opposite circulations.
For rotors of equal vorticity Eq.~(\ref{3.4})  for the center of mass coordinate reduces to $\partial_t{\bf R}=0$, indicating that in this  case the position of the center of mass of the rotor pair remains fixed at its initial value.
We now discuss the dynamics of various pairs of active and passive rotors. 
\newcommand{\w}{0.32}
\begin{figure}[h]
\centering
\begin{tabular}{p{\w\columnwidth}p{\w\columnwidth}p{\w\columnwidth}}
\centering (a) & \centering (b) & \centering (c)
\end{tabular}
\includegraphics[width=0.32\linewidth]{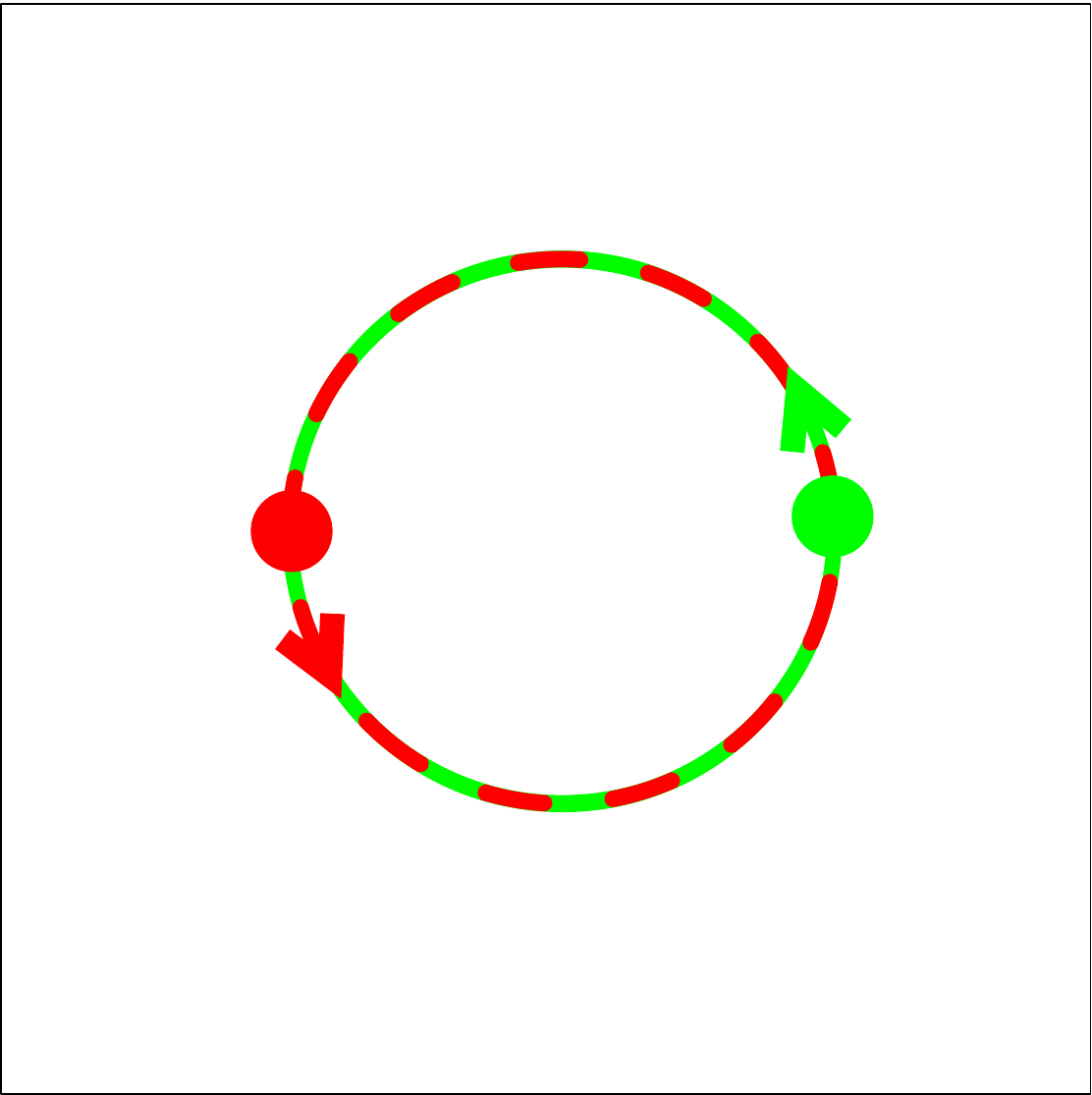}
\includegraphics[width=0.32\linewidth]{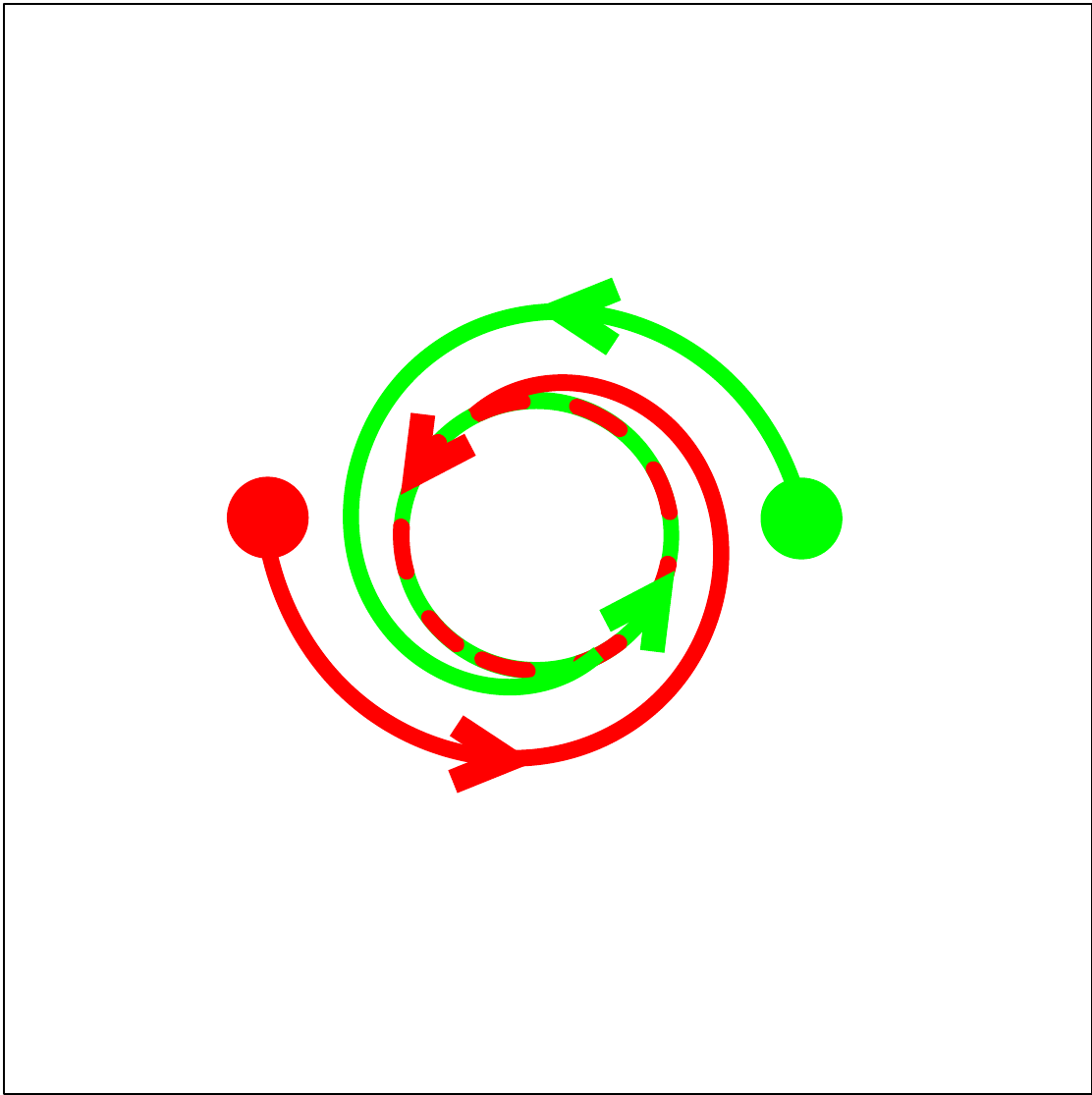}
\includegraphics[width=0.32\linewidth]{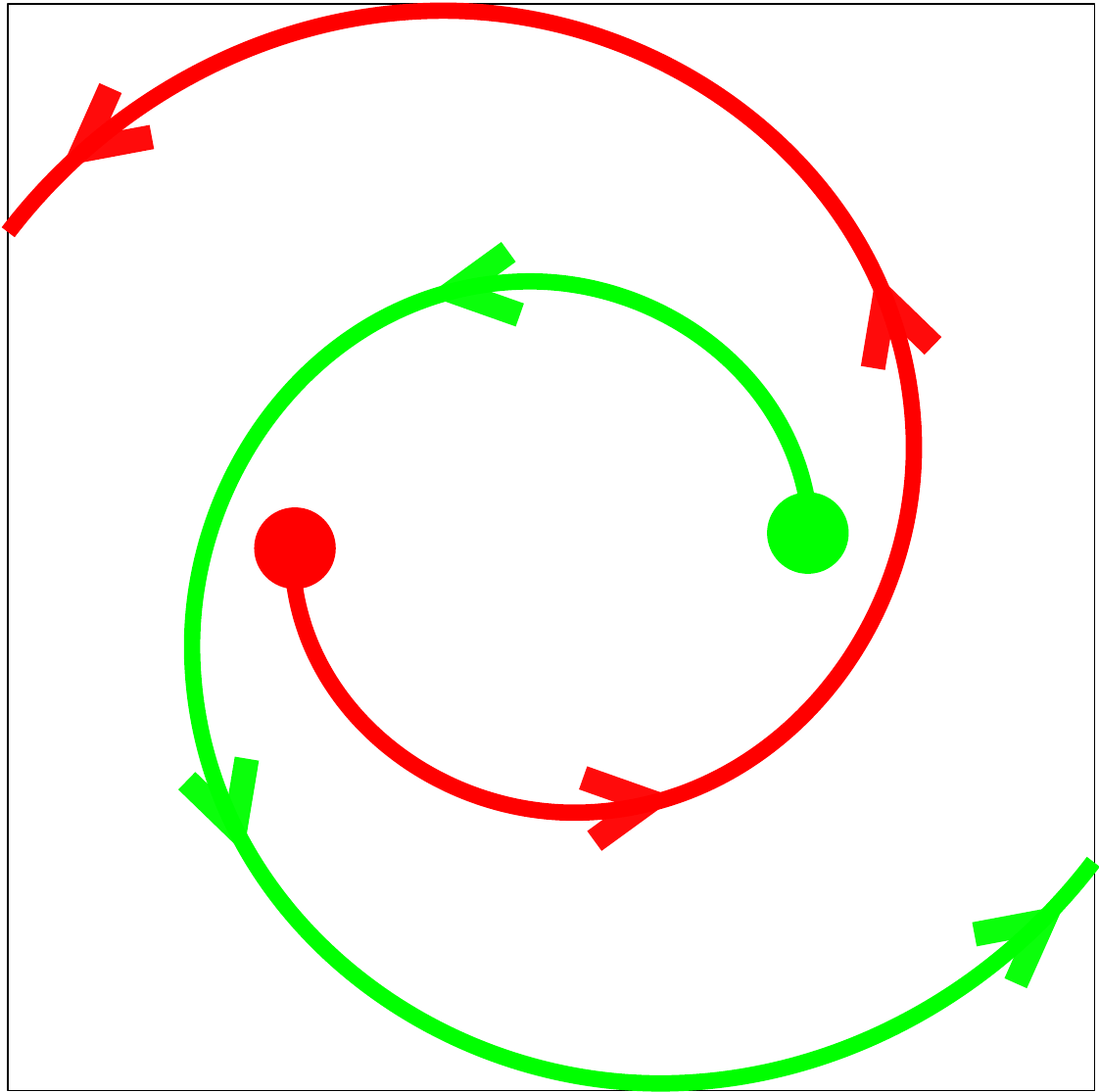}
\caption{Schematic trajectories of two identical rotors of equal vorticity.
The two dots are the initial positions of the rotors.
A line with alternating color is used where the trajectories are superimposed.
(a) Pair of passive rotor or pure active rotor (circle).
(b) Pair of contractile swimmers (shrinking spiral).
(c) Pair of tensile swimmers (expanding spiral).}
\label{Trajectories1}
\end{figure}
\begin{enumerate}
\item {\em Like passive rotors.} In this case $\epsilon=1$ and the center of mass is static. The radial component of the force reduces to the short-range repulsion force. The two rotors rotate about their center of mass with an angular velocity
$\Omega =\tau^e/(4\pi\eta r_0^3)$
describing a circle of diameter $r_0$.
The diameter $r_0$ is simply the initial separation of the rotors if
they start out at a separation greater than the range $d$ of the repulsive
interaction and equals $d$ if the initial separation of the centers of the two rotors is less than $d$. The
trajectory  is as shown in Fig.~(\ref{Trajectories1}a).

\item {\em Like active rotors.} Again $\epsilon=1$ and the center of mass of the pair remains fixed. The rotors experience active radial forces $F^a$ that can be  attractive ($0\leq\vert\phi\vert<\pi/2$, corresponding to pullers) or repulsive ($\pi/2<\vert\phi\vert\leq\pi$, corresponding to  pushers).
The relative coordinate of the pair of swimmers describes a circular trajectory of varying radius, corresponding to an inward bounded spiral (Fig.~\ref{Trajectories1}b) or an outward unbounded spiral (Fig.~\ref{Trajectories1}c),
depending on the nature of the swimmers. Writing ${\bf r}$ in polar coordinates as  ${\bf r}=(r,\theta)$
we can solve for  the dynamics of the angular coordinate, which is given by
\[
\theta(t) =\frac{3}{8}\ell^2\tan\phi \left[ \frac{1}{r_0^2}
-\frac{1}{\left(r_0^{3}+\frac{3F\ell\cos\phi }{8\pi\eta}t\right)^{2/3} }
\right]\;.
\]%
For tensile swimmers (pushers) the angular dynamics slows down as $t^{-1/3}$ in the
asymptotic regime, while for contractile swimmers (pullers), the dynamics is
regularized by the onset of the short-range repulsion. Then $\theta(t)\sim t$ at long times and the two rotors eventually settle into circling each other describing a circle
of diameter $r_e$ where $r_e<d$ is such that the attractive force is balanced by the short-range repulsive one.
In the limit case when $\phi=\pi/2$ (corresponding to pure rotors), the radial interaction reduces to the short-range repulsion and the dynamics is similar to that of a pair of like passive rotors with an angular velocity $\Omega$ now equal to $A^a/(2r)\sim1/r^5$ instead of $A^p/(2r)\sim1/r^3$.
\item {\em Unlike passive rotors.} In this case $\epsilon=-1$ and the
dynamics of the relative coordinate becomes purely radial and, because the rotors are passive, repulsive and short-ranged.
The center of mass coordinate acquires a velocity in the
direction orthogonal to the line joining the two rotors. The 
pair of rotors becomes "self-propelled", in that it develops sustained motion in the
direction orthogonal to the line joining their centers. For initial separation $r_0>d$, the two rotors
move along parallel lines at a speed 
$V=\tau^e/(8\pi \eta r_0^2)$, as shown in Fig.~\ref{Trajectories2}a.
For initial separations $r_0<d$ there is an
initial transient associated with the two rotors moving apart due to the
repulsive force, after which they settle into the self-propelled trajectory along parallel lines
 shown in Fig.~\ref{Trajectories2}a with a speed 
$V=\tau^e/(8\pi \eta d^2)$.

\item {\em Unlike active rotors.} As in the case of like rotors, the trajectory of unlike active rotors is similar to that of passive ones,
with additional dynamics that arises from the repulsive/attractive radial part of the
hydrodynamic interaction. For the case of attractive interactions, \emph{i.e.},
pullers, there is an initial transient during which the two rotors
approach each other until the attractive interaction is balanced by the short-range repulsion.
After this time, the two rotors move as a self-propelled pair along parallel
straight lines with a velocity that scales as the inverse fourth power of their
separation (see Fig.~\ref{Trajectories2}b). To understand the behavior of pushers it is useful to explicitly integrate the equation of motion to obtain the time evolution of the magnitude of the center of mass coordinate, given by%
\[
R(t)= \frac{3}{8}\ell^2\tan\phi 
\left[ \frac{1}{r_0}-\frac{1}{%
\left(r_0^3+\frac{3F\ell\cos\phi}{8\pi\eta}t\right)^{1/3}} 
\right]\;.
\]%
It is then evident that the  "self-propulsion" speed of the pair decays with
time. The active force driving the
self-propulsion  scales as $1/r^{4}$ while the  force driving the
repulsive separation between the rotors scales as $\frac{1}{r^{2}}.$ As a result,
the repulsive interaction eventually wins and the trajectories
of the two rotors diverge  from each other, as shown in  Fig.~\ref{Trajectories2}c.
As for like rotors, pure active unlike rotors ($\phi=\pi/2$) behave like passive unlike rotors, but move with a speed $V\sim1/r^4$ instead of $1/r^2$.
\end{enumerate}
\begin{figure}[h]
\centering
\begin{tabular}{p{\w\columnwidth}p{\w\columnwidth}p{\w\columnwidth}}
\centering (a) & \centering (b) & \centering (c)
\end{tabular}
\includegraphics[width=0.32\linewidth]{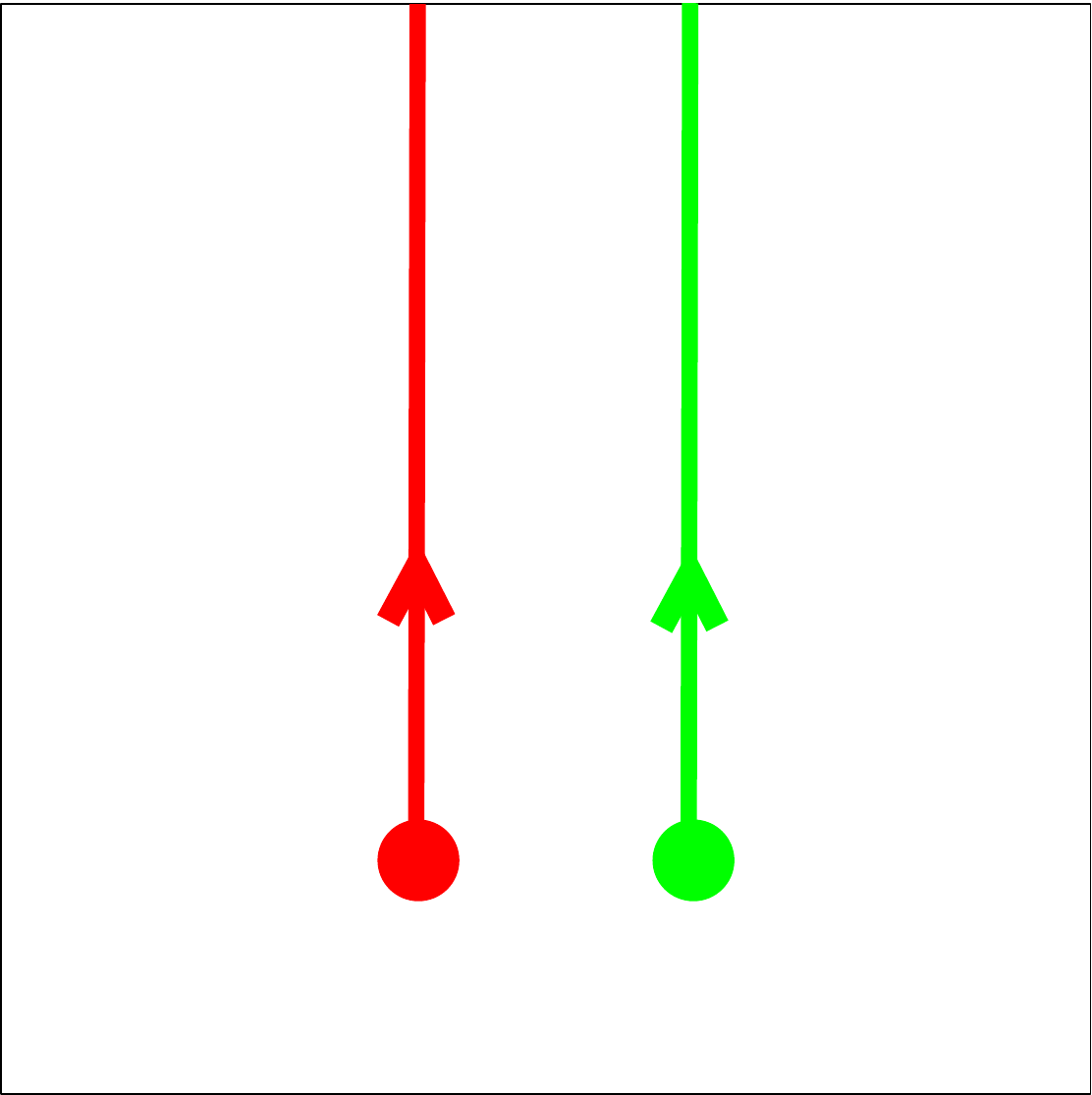}
\includegraphics[width=0.32\linewidth]{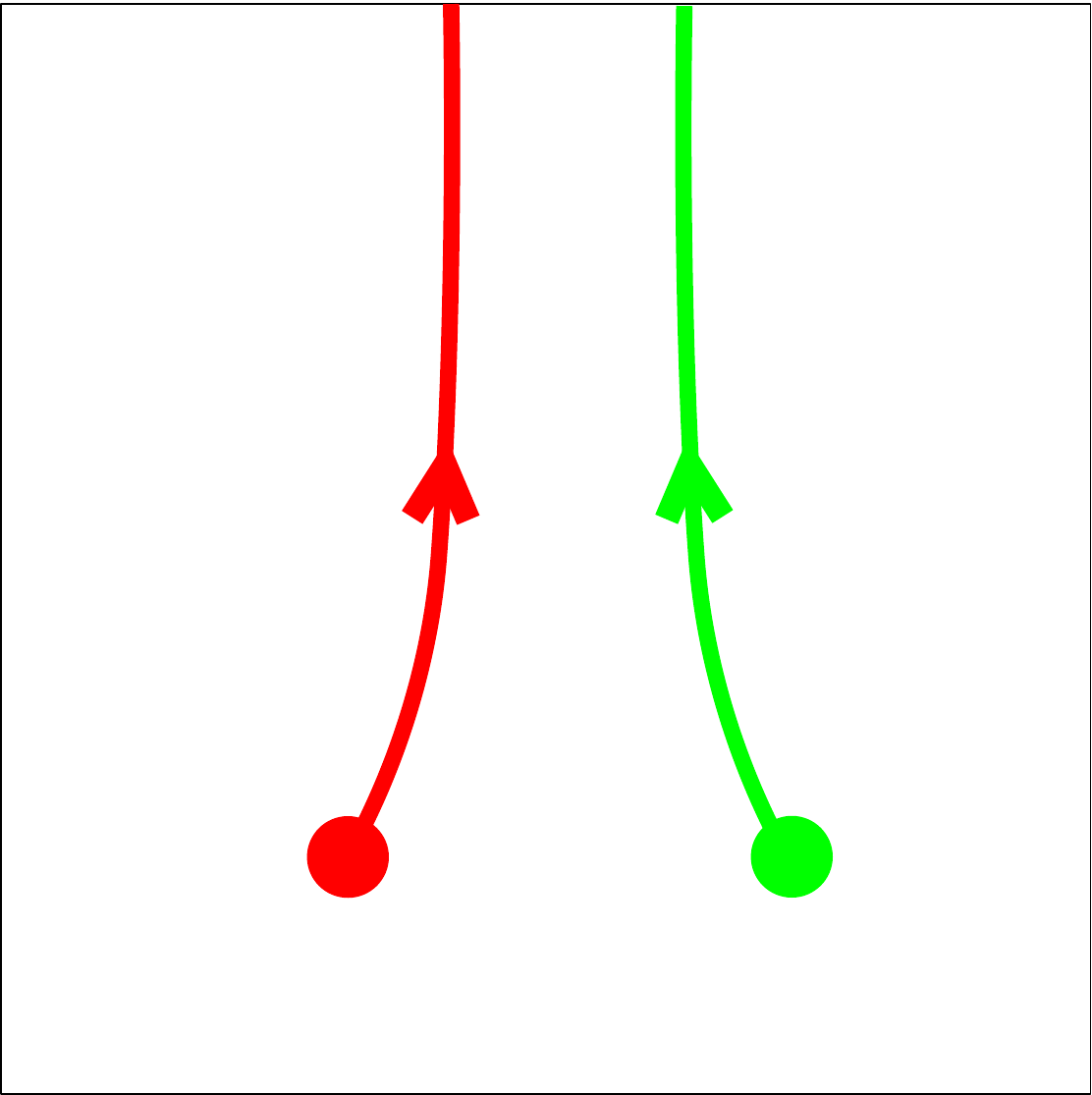}
\includegraphics[width=0.32\linewidth]{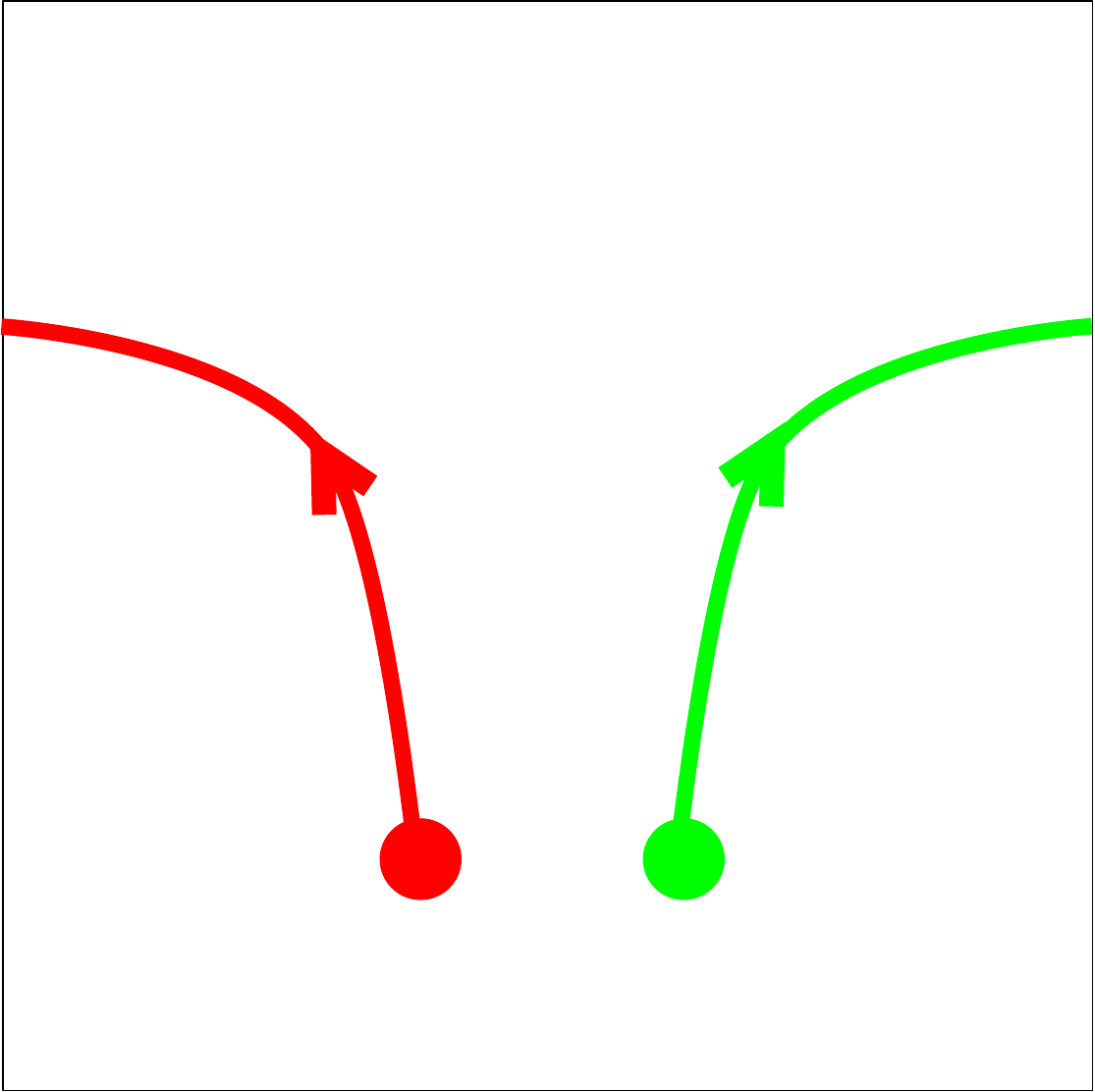}
\caption{Schematic trajectories of two identical rotors of opposite vorticity.
The two dots are the initial positions of the rotors.
(a) Pair of passive rotors or pure active rotors (straight parallel lines).
(b) Pair of contractile swimmers (converging lines).
(c) Pair of tensile swimmers (diverging lines). }
\label{Trajectories2}
\end{figure}

Summarizing, like rotors circle around each other, while their center of mass remains stationary. Unlike rotors are propelled in the direction orthogonal
to the line joining their centers. The difference between passive rotors and
active rotors primarily arises from the presence of active  radial
hydrodynamic interactions that can be attractive or repulsive depending on
the nature of the swimming mechanism. 
When the active propulsion mechanism
results in pure rotation (\emph{i.e.}, when $\phi =\pi /2$), the
difference between active and passive rotor trajectories lies only in the
scaling of the velocities as a function of rotor separation.

\section{Discussion}
\label{discussion}

The important role of the azimuthal hydrodynamic interactions between colloidal particles in a fluid driven to rotate by an external drive in controlling the dynamical assembly of such artificial microswimmers has been recognized for some time~\cite{Grzybowski2000,Grzybowski2001,Climent2007,Barbic2001,Schreiber2010}. 
Yet, only recently it has been pointed out that this interaction yields a net finite velocity for the center of mass of two rotors of opposite vorticity, which will therefore behave as a self-propelled pair~\cite{Leoni2010a}. Such a ``cooperative self-propulsion'' is a novel effect with potential applications.
Experimental realizations of passive rotors naturally generate, however, rotors of the same vorticity. The most common realization of collections of microrotors consist of magnetic dipoles driven by an  external rotating magnetic field~\cite{Grzybowski2000,Grzybowski2001,Climent2007}.
The direction of rotation of the rotors is then imposed by that of the magnetic field, and is the same for all the rotors.
In Ref.~\citeN{Barbic2001}, the drive is applied in a more local way by using a small coil that is positioned right above the rotor.
Using two such coils would allow to apply opposite drives on two rotors.
Similarly, a laser trap can be used to control the rotation of an individual rotor~\cite{Friese1998,Friese1998a}.
Still, observing self propulsion with either of these two devices would require that the sources of the angular drive follow the rotors in their motion.

Internally driven rotors, on the other hand, do not suffer from this limitation and may be a good candidate for the observation of cooperative self propulsion.
In these active systems, however, the internally generated torque that is transmitted by the swimmers to the fluid is exactly balanced by the torques transmitted by the fluid to the swimmers. As a result, the azimuthal component of the hydrodynamic interaction vanishes to dipolar order. We show that such an azimuthal component of the interaction among active rotors arises nevertheless to octupolar order and remains finite upon cycle-averaging. While the dipolar azimuthal interaction among passive swimmers decays as $1/r^2$, the corresponding interaction among active rotors decays as $1/r^4$.

\changeA{
\changeD{
It is important to stress that the following results apply for a general active rotor, regardless of its shape :}
\\ \indent
(i)~The dipolar flow field has no azimuthal component.
This result is not new and has been known for a long time~\cite{Batchelor1970}. 
It is due to the torque-free nature of the swimmers and the fact that the \emph{rotlet}/\emph{stresslet} decomposition coincides with the radial/azimuthal one in the plane orthogonal to the axis of rotation.
\\ \indent
(ii)~The quadrupolar flow vanishes when averaged over a period of rotation.
This relies on \changeD{the two-dimensional nature of the problem and} the equivalence between the period average and an orientation average, which holds if the period of rotation is much smaller than any other time scale in the problem. The vanishing of the cycle-averaged quadrupolar contribution then follows immediately from the symmetry properties of the multipole expansion of the flow field.
\\ \indent
(iii)~The octupolar flow field remains finite upon cycle averaging and results in an azimuthal interaction force that decays as $1/r^4$.
}

Furthermore, these results can be extended to a more general form of the Green's function.
We have limited ourselves to a discussion of the flow field in planes normal to the rotor's axis of rotation. A motivation for this is that rotors are often confined to a liquid/air, liquid/liquid or liquid/solid interface. On the other hand, such confinement often changes the form of the Green's function.
Assuming that the medium is isotropic in the plane perpendicular to the rotors' axis of rotation, and to which they are confined, the most general Green's function describing the flow field in this plane can be written as 
${\cal O}_{ij}({\bf r})=A(r)\ \delta_{ij} +B(r)\ \hat{r}_i\hat{r}_j$ where $A$ and $B$ are two unknown function of $r=|{\bf r}|$.
In addition, we assume that at distances large compared to the thickness of the interface ${\cal O}_{ij}({\bf r})$ decays as $1/r^n$, 
\emph{i.e.} ${\cal O}_{ij}({\bf r})=(a\ \delta_{ij} +b\ \hat{r}_i\hat{r}_j)/r^n$ where $a$ and $b$ are two unknown constants.
The three-dimensional Oseen tensor used in this work  corresponds to $n=1$ and $a=b=1$.
The Green's function describing interactions in a thin fluid film of two-dimensional viscosity $\eta$, surrounded by by a bulk  fluid of viscosity $\eta_e$  corresponds to $n=1$, $a=0$ and $b=2\eta/\eta_e$~\cite{Leoni2010a}.
A higher value of $n$ can be found, \emph{e.g.}, if the rotors are
close to a solid wall with no-slip boundary condition. The $1/r$ term is then canceled by the presence of a virtual source of opposite strength located on the other side of the wall, leading to $n=2$~\cite{Hasimoto1980,Ramia1993}.
The flow field resulting from any of these Green's functions is qualitatively the same as the one derived in this work.
More specifically, the points (i), (ii) and (iii) above remain true regardless of the values of $a$, $b$ and $n$, \changeD{except that the octupolar term now decays as $1/r^{n+3}$}~$^*$.
\footnotetext{*~%
\changeA{%
Note that there is a special case, namely $na+b=0$, in which there is no azimuthal field at all, regardless of the nature of the rotor.}
}

\changeA{
On the other hand, the exact form of the azimuthal octupolar interaction between active rotors depends on the nature of the rotor.
The one-bead swimmer we described in this paper is a simple model system in that
it has an isotropic drag (via the spherical bead) and rotates around itself.
A more general model would feature \changeE{an} anisotropic drag (e.g. because of a non-spherical bead or \changeE{the presence of} more than one bead) and/or a significant unperturbed trajectory radius $R_0$. The azimuthal interaction between two rotors would then depend on their relative orientation, the dynamics of which would have to be worked out.
This is beyond the scope of the present paper and left for future work.
}

\appendix

\makeatletter 
\renewcommand*{\@seccntformat}[1]{Appendix \csname the#1\endcsname:~}
\makeatother 

\section{Period averaged flow field}
\label{period_avg}

The flow field generated by a rotor depends in general on its orientation. As the particle rotates and moves through the fluid, the flow field evolves in time as well.
This evolution is characterized by two well separated time scales: the period of rotation $T_0$ of the rotor and the (generally much longer) time scale associated with its motion through the fluid due to interactions with other rotors.
On time scales large compared to $T_0$, the swimmers' dynamics is governed by an effective flow field obtained by averaging the instantaneous flow field over a period of rotation.
More precisely, the corrections  to the average flow field $u$ arising from instantaneous fluctuations over time scales $T_0$ are of order $\delta u \sim T_0 u \nabla u\sim u^2 T_0/r$ where $T_0 u$ is the typical distance travelled by the rotor during the time $T_0$ by following the flow field of other rotors located at distance $r$.
The leading term of the flow field is at most dipolar, with $u \sim 1/r^2$.
The correction is then $\delta u \sim 1/r^5$ and is negligible compared to terms up to octupole order ($\sim 1/r^4$) in the average flow field.

Similarly, the fluctuations in the angular velocity are governed by the ratio of the curl of the flow field in Eq.~\eqref{drag_T}, which is at most of order $F/\eta r^3$ (corresponding to a dipolar flow),
to the angular velocity $\omega \sim F\ell/\zeta^\textsc{r}$ given in Eq.~\eqref{dynO_swim1}.
As a result, we can assume that the rotation of an individual swimmer takes place at a constant speed
(up to corrections of order $1/r^5$) and replace the average over a rotation by an average over the orientation of the swimmer.

These observations greatly simplify the dynamics. In particular,
 the quadrupole term in the multipole expansion of the flow field, which changes sign when the rotor is rotated by $\pi$, does not contribute to the averaged flow.
It should be stressed, however, that the angular averages washes out the angular dependance of the flow field,
yielding an isotropic effective flow.

Finally, a study of the corrections due to these fluctuations on the dynamics of passive rotors can be found in Refs.~\citeN{Leoni2010a,Schreiber2010}.

\section*{Acknowledgments}
The work at Syracuse was supported by the NSF on grants DMR-0806511 and DMR-1004789. MCM thanks Tannie Liverpool and Marco Leoni for illuminating discussions.
YF thanks Marco Polin for discussions about swimming microorganisms.
AB acknowledges support from the Brandeis MRSEC (DMR-0820492)

\footnotesize{
\bibliography{biblio}
\bibliographystyle{rsc}
}

\end{document}